\documentclass[useAMS,usenatbib,usegraphicx]{mn2e}
\usepackage{amssymb}

\usepackage{times} 

\newcommand{\degree}{^\circ}

\def \<{\langle}
\def \>{\rangle}

\title[Observation number correlation in WMAP data]
{Observation number correlation in WMAP data}

\author[T. P. Li et al.]{Ti-Pei Li$^{1,2}$\thanks{E-mail: litp@tsinghua.edu.cn},
Hao Liu$^{1}$\thanks{E-mail: liuhao@ihep.ac.cn}, Li-Ming Song$^{1}$, Shao-Lin Xiong$^{1,3}$ and
Jian-Yin Nie$^1$\\
$^{1}$Key Laboratory of Particle Astrophysics, Institute of High Energy Physics,
Chinese Academy of Sciences, Beijing, China\\
$^2$Center for Astrophysics, Tsinghua University, Beijing, China\\
$^3$Graduate School of Chinese Academy of Sciences, Beijing, China
}
\begin{document}

\date{}

\pagerange{\pageref{firstpage}--\pageref{lastpage}} \pubyear{2009}

\maketitle

\label{firstpage}

\begin{abstract}
A remarkable similarity between the large-scale non-Gaussian pattern of cosmic microwave
background (CMB) temperatures obtained by Wilkinson Microwave Anisotropy Probe (WMAP)
mission and the distribution feature of observation numbers is noted.
Motivated from such a similarity, in this work we check the WMAP data for
the correlation between pixel temperature $t$  and observation number $N$.
Systematic effect of imbalance differential observation and significant
$t$-$N$ correlation in magnitude, distribution non-Gaussianity and
north-south asymmetry are found. Our results indicate that, for precision
cosmology study based on WMAP observations, the observation
effect on released WMAP temperature maps has to be further carefully studied.
\end{abstract}


\section{INTRODUCTION}
The WMAP observations provide precision data for cosmology study.
 By analyzing CMB maps from the first year WMAP (WMAP1) data,
Tegmark et al. (2003) find both the CMB quadrupole and octopole having
 power along a particular spatial axis and more works \cite{cos04,eri04a,sch04,jaf05}
find that the axis of maximum asymmetry tends to lie close to the ecliptic axis.
A similar anomaly was also found in COBE maps \cite{cop06}.
The unexplained orientation of large-scale
patterns of CMB maps in respect to the ecliptic frame is one of the biggest surprises
in CMB studies \cite{sta05}.
A notable asymmetry of temperature fluctuation power in two opposing hemispheres is also found
in the WMAP1 and COBE results  \cite{eri04b, han04}.
After the release of more WMAP results, similar large-scale anomalies
are still detected
in the WMAP3 data \cite{abr06,jaf06,cop07,lan07,eri07,par07,vie07,sam08}
and WMAP5 data \cite{ber08} as well.

These apparent anomalies, if found to be cosmological origin, will pose a big challenge
to the standard model of cosmology.
Therefore, inspecting the effects of WMAP observation on released data
at large angular scales more carefully is worth doing.
We show in \S2 that there exists in WMAP data a remarkable similarity
between the large-scale non-Gaussian pattern of map temperatures
and the distribution feature of observation numbers.
Motivated from such a similarity, in this work we further check the WMAP data
for the correlation between pixel temperature $t$  and observation number $N$.
A systematic effect of imbalance differential observation and
significant $t$-$N$ correlation in magnitude, distribution non-Gaussianity and
north-south asymmetry  are detected and shown in \S3.
We give a brief discussion on
the observation effect in WMAP data in \S5.

\section{LARGE-SCALE NON-GAUSSIAN MODULATION}

 \begin{figure}
   \vspace{2mm}
\includegraphics[width=0.271\textwidth,angle=90]{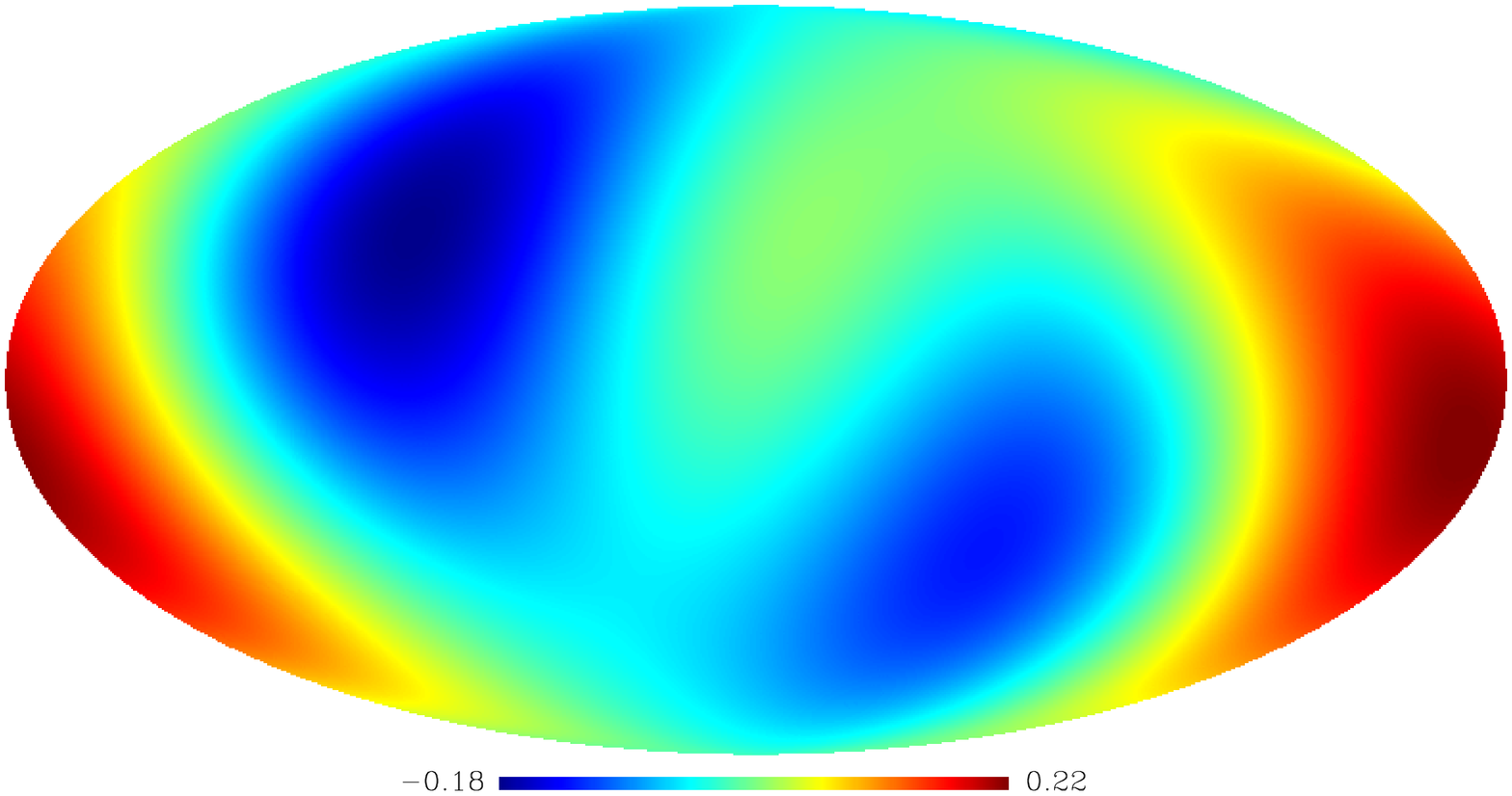}
\includegraphics[width=0.271\textwidth,angle=90]{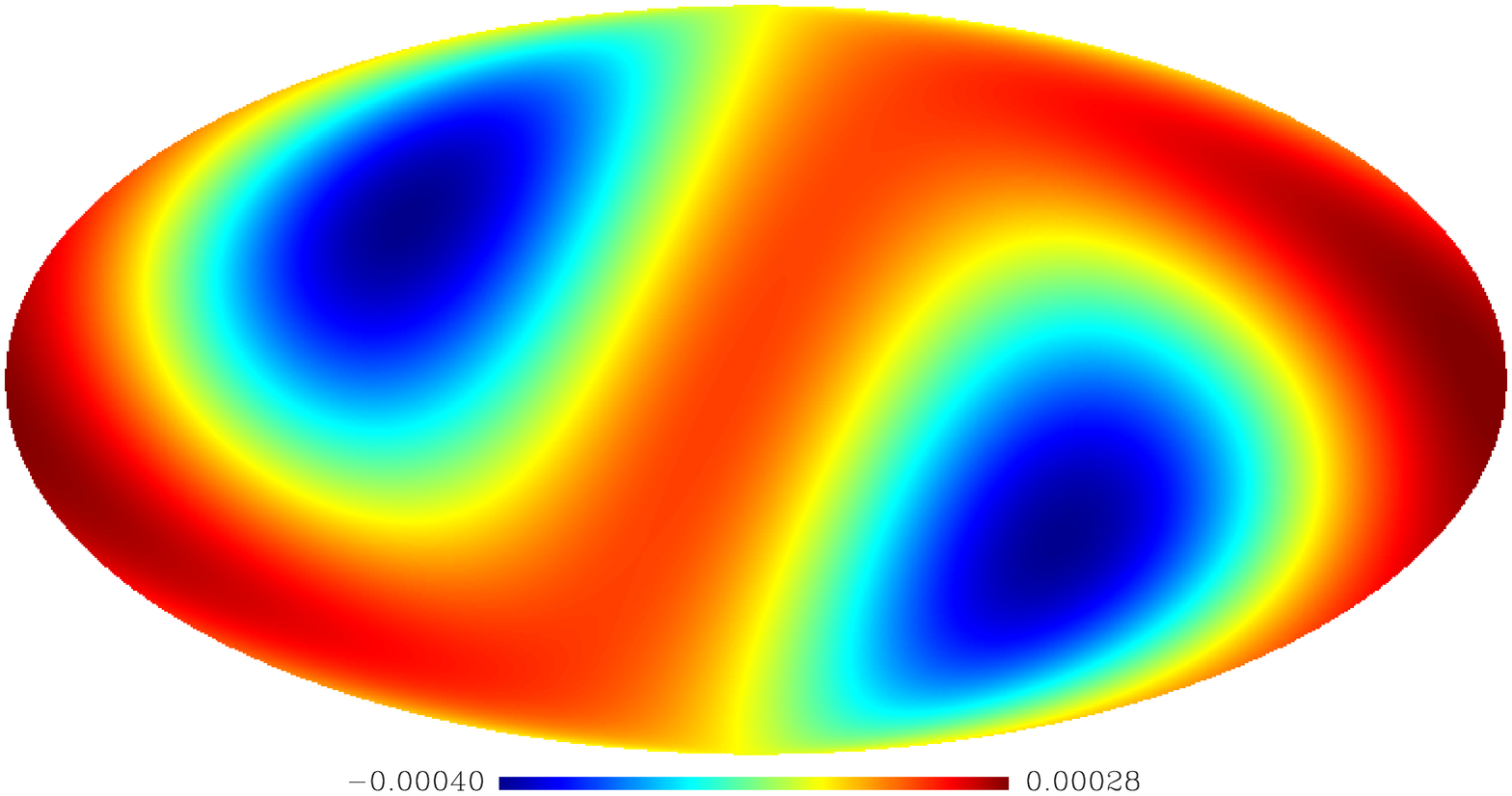}
\includegraphics[width=0.271\textwidth,angle=90]{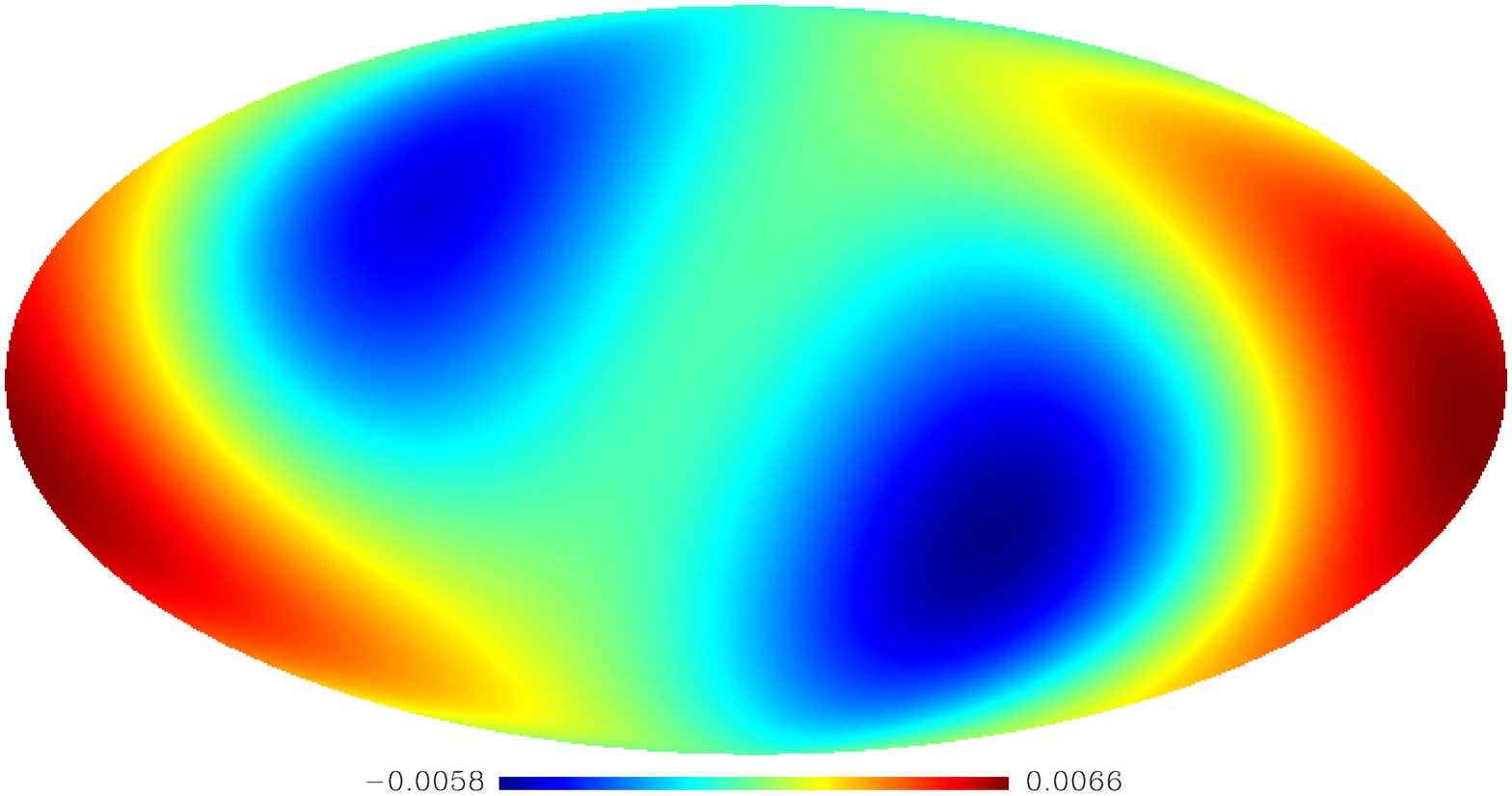}
\caption{Large-scale non-Gaussian modulation features with $l_{max}=2$
for the WMAP3 V-band data .
{\sl Top panel}: Best-fit large-scale modulation function $f(\mathbf{\hat{n}})$
for temperature map.
 {\sl Middle panel}: large-scale feature of $1/N$ map.
{\sl Bottom panel}: large-scale feature of $1/\sigma(N)$ map.}
  \label{modu}
\end{figure}

To address the large scale anomalies, such as asymmetry,
alignment and low $l$ power issues detected in WMAP data
with different techniques, the WMAP team~\cite{sper06} describe the observed
temperature fluctuations, $\mathbf{{\hat{t}}}$, as a Gaussian and isotropic random
field,  $\mathbf{t}$, modulated by a function $f(\mathbf{n})$
\[ \hat{t}(\mathbf{n})=t(\mathbf{n})[1+f(\mathbf{n})] \]
where $f(\mathbf{n})$ is an arbitrary modulation function. They expand $f(\mathbf{n})$
in spherical harmonics
\[ f(\mathbf{n})=\sum_{l=1}^{l_{max}}\sum_{l=-m}^mf_{lm}Y_{lm}(\mathbf{n}) \]
and use maximum likelihood technique with
a Markov Chain Monte Carlo solver to get the best fit values of $f_{lm}$
with $l_{max}=2$ for the WMAP3 V-band map.
 The top panel of Fig.~\ref{modu} is obtained based on the best fit coefficients, showing
in a unifying manner the large scale anomalies in WMAP temperature fluctuations
which is the same feature that has been identified in a number
of papers on non-Gaussianity.

The sky coverage of WMAP mission is inhomogeneous
-- the number of observations being greatest at the ecliptic poles
and the ecliptic plane being most sparsely observed~\cite{hin07}.
To address the observation number distribution in the same way
for the temperature distribution,
we calculate the spherical harmonic coefficients $f_{lm}$ with $l_{max}=2$
for the map of $1/N$ with $N$ being number of observations per sky pixel
from the WMAP3 V-band data. The middle panel of Fig.~1 shows
the map of $1/N$ reconstructed based on the coefficients $f_{lm}$.
Comparing the two graphs at the top and middle of Fig.~1,
we can see that the large-scale non-Gaussian modulation features of WMAP temperature map
and scan pattern being similar for each other, showing considerable connection
between pixel temperature $t$ and observation number $N$.

We further show in the bottom panel the reconstructed result for
the observation fluctuation map --
the map of $1/\sigma(N)$ where the rms variation
$\sigma(N)=\sqrt{\<(N-\<N\>)^2\>}$
calculated within a region of $\sim 1\degree$ side dimension for each sky pixel.
From this map we notice, unexpectedly to some extent,
that observation numbers in released WMAP data are also fluctuated.
 In comparing the three panels in  Fig.~\ref{modu},
the modulation pattern for the observation fluctuation map (the bottom panel)
is more similar to the detected anomalies (the top panel),
indicating that the fluctuation of observation numbers could produce additional
uncertainty to the recovered temperature map.

\section{TEMPERATURE-EXPOSURE COUPLING}
\subsection{$t$-$N$ Correlation}
The remarkable similarity between the large-scale non-Gaussian pattern of WMAP CMB
temperatures and the distribution feature of observation numbers shown in the previous section
prompts us to check the WMAP data for the correlation
between pixel temperature $t$  and observation number $N$.
To inspect the $t$-$N$ coupling, for a sky pixel $i$ we calculate the correlation coefficient
$C_{_{t-N}}(i)$  by
\begin{equation}\label{c}
C_{_{t-N}}(i)=\frac{\sum_j(t_i(j)-\overline{t_i})(N_i(j)-\overline{N_i})}
{\sqrt{\sum_j(t_i(j)-\overline{t_i})^2 \sum_j(N_i(j)-\overline{N_i})^2}}\,,
\end{equation}
where the summations are taken over all WMAP pixels $j$ (in the original resolution) 
within a spherical cap centered at the vertex $i$ with an angular radius of $10\degree$. 
We use Eq.~\ref{c} to produce a correlation
map $C_{_{t-N}}(i)$ from the WMAP5 Q-, V- and W-band data separately, 
where the vertexes $i$ are defined with $r5$ resolution of HEALPix pixelization 
scheme \cite{gor05} in the sky sphere, if more than $20\%$ of pixels
of a cap are inside the Galactic mask KQ85 \cite{gol08,nol08}, the cap is no longer used. 
From a correlation map, we calculate the average of absolute correlation 
coefficients $\<|C_{_{t-N}}|\>$ over the full sky,
$\<|C_{_{t-N}}|\>_{south}$ over the South Galactic hemisphere, 
$\<|C_{_{T-N}}|\>_{north}$ over the north
Galactic hemisphere, and the south-north asymmetry ratio 
$\<|C_{_{t-N}}|\>_{south}/\<|C_{_{t-N}}|\>_{north}$.

Now we use simulations to test the significance of the magnitude of
$t-N$ correlation and its north-south asymmetry in WMAP data. The
program synfast in HEALPix software package (available at
http://healpix.jpl.nasa.gov) can create temperature maps computed as
realizations of random Gaussian fields on a sphere characterized by
the user provided spherical harmonic coefficients of an angular
power spectrum. For each studied band, we produce 50,000 simulated
temperature maps with the synfast program from the best fit
$\Lambda$CDM model power spectrum \cite{nol08} with the beam
function \cite{hil08} and five-year like noise \cite{lim03}. We
compute the average absolute value of $t$-$N$ correlation
coefficient and its north-south asymmetry ratio for each simulated
CMB map in the same way as for the WMAP5 data. Finally, we calculate
the average and its standard deviation from 50,000
$\<|C_{_{t-N}}|\>$ and 50,000
$\<|C_{_{t-N}}|\>_{south}/\<|C_{_{t-N}}|\>_{north}$ respectively.

The obtained results are summarized in Table~\ref{tabTN}. From
Table~\ref{tabTN} we can estimate the significance of the average
absolute magnitude of $t-N$ correlation being $4.68\sigma$,
$4.23\sigma$ and $4.78\sigma$ for Q-, V- and W-band respectively.
Therefore, the $t-N$ coupling in WMAP5 data is much stronger than
what expected from the $\Lambda$CDM model. In other words, the
inhomogeneity of WMAP exposure may produce notable distortion in
temperatures observed for the $\Lambda$CDM CMB. Especially, for each
of the three bands, none of the 50,000 estimators $\<|C_{_{t-N}}|\>$
obtained from simulation exceeds the WMAP values in Table. 1. This
is well consistent with the $\sim4\sigma$ significance estimation
for the three bands. From Table~\ref{tabTN} we can also see a
considerable north-south asymmetry existing in $t-N$ correlation
with significance of $2.58\sigma$, $2.18\sigma$ and $2.21\sigma$ for
Q-, V- and W-band respectively.

\begin{table*}
\begin{minipage}{100mm}
\caption{Average $t$-$N$ correlation coefficients \label{tabTN}}
 \begin{tabular}{cccc}
\hline
\hline
 & & $\<|C_{_{t-N}}|\>$ & $\frac{\<|C_{_{t-N}}|\>_{south}}{\<|C_{_{t-N}}|\>_{north}}$\\
\hline
Q-band & WMAP5       & 0.116 & 1.25 \\
      & Expectation  & $0.0837\pm 0.0069$ & $1.010\pm0.093$ \\
\hline
V-band &WMAP5       & 0.0977 & 1.20 \\
     & Expectation  & $0.0719\pm 0.0061$ & $0.993\pm0.095 $   \\
\hline
W-band&WMAP5       & 0.0942 & 1.21\\
      &Expectation  & $0.0660\pm 0.0059$ & $0.995\pm0.097$   \\
\hline
\end{tabular}
\end{minipage}
\end{table*}

\begin{figure}
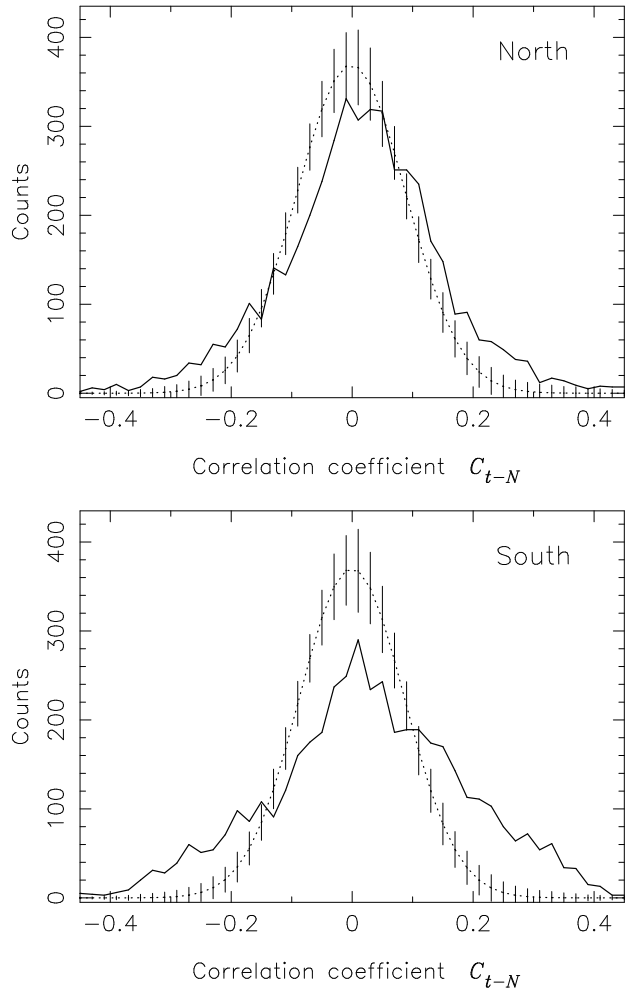

\includegraphics[width=0.36\textwidth,angle=270]{f2a.ps}\vspace{3mm}
\includegraphics[width=0.36\textwidth,angle=270]{f2b.ps}\vspace{3mm}
 \vspace{2mm}  \caption{Histograms of $t$-$N$ correlation coefficients.
{\it Solid line}: from the WMAP5 Q-band data.
{\it Dotted line}:  expectation and $1\sigma$ error bar from 50000 simulations
for the $\Lambda$CDM CMB.
{\sl Upper panel}: north hemisphere.
{\sl Lower panel}: south hemisphere.
}
   \label{fhist}
   \end{figure}

Furthermore, we test the Gaussianity of obtained WMAP $t$-$N$ correlation distribution.
Fig.~\ref{fhist} shows histograms of $t$-$N$ correlation coefficients from WMAP5 Q-band data
(solid line) and simulations (crosses and dotted line) for the Galactic north
and south hemispheres separately.
From Fig.~\ref{fhist} we see that, for both northern and southern hemispheres,
the simulated $t$-$N$ correlation coefficients are normally distributed,
and the correlation coefficients measured from WMAP5 data visibly deviate
from the normal distribution.
The histogram of $t$-$N$ correlation for the Galactic south hemisphere from WMAP5 Q-band data
is much higher than the error bars at both tails, indicates that there are far more caps
having strong $t$-$N$ correlation  than expected. On the contrary,
the histogram for the north hemisphere is rather close to
the simulation result. These results on non-Gaussianity and
north-south asymmetry for $t$-$N$ correlation distribution are consistent with
that for $t$-$N$ correlation magnitude.

The Kolmogorov-Smirnov test is also used to compare the measured ($F$) and simulated
($F_0$) cumulative distribution functions (CDFs). 
We find the maximum absolute differences ($D_n$)
between the measured and simulated CDFs, $D_n=max(|F-F_0|)$, is  
 0.105 for the north hemisphere and 0.185 for the
south hemisphere, respectively.
The threshold of $D_n$ for null hypothesis $F=F_0$ at
$\alpha=0.01$ significance level and sample size, the used sky pixels $N_p\sim 4400$,
 is known to be $1.63/\sqrt{N_p}\approx 0.024$. The measured $D_n$
are much higher than the threshold. Therefore, the null hypothesis $F=F_0$ has been
rejected at at $\alpha=0.01$ or even higher significance level, 
indicating that the measured $t$-$N$ correlation deviates significantly from simulation.

\subsection{Effect of Imbalance Differential Observation}
We find that the above results from the WMAP5 foreground-cleaned Q-band data
with the KQ85 mask are almost the
same with what obtained from the data without any mask; therefore,
the detected $t$-$N$ correlation can not be explained by the foreground effect.
We will show that the instrumentation and observation effect can contribute
to the observed $t$-$N$ correlation and then distort observed temperature maps.

\subsubsection{Differential observation}
The COBE and WMAP missions measure
temperature differences between sky points
using differential radiometers consisting of plus-horn and minus-horn
 with a fixed separation angle $\theta_{beam}$~\cite{smo90,ben03a}.
The beam separation angle of WMAP radiometers is $\theta_{beam}\sim 141\degree$.
Let denote $t_i$ the temperature anisotropy at a sky pixel $i$.
The raw data in a certain band is a set of
temperature differences {\bf d} between pixels in the sky. From $N$ observations we have
the following observation equations
\[
\begin{array}{c@{\:-\:}c@{\;=\;}c}
t_{1^+} & t_{1^-} & d_1 \\
t_{2^+} & t_{2^-} & d_2 \\
\multicolumn{3}{c}{\dotfill}\\
t_{N^+} & t_{N^-} &~~d_N~,
\end{array}
\]
or in matrix notation
\begin{equation}
\label{dt1}
\mathbf{At=d}~.
\end{equation}
Where the scan matrix  {\bf A}$ =(a(k,i)),~k=1,\cdots,N$
and $i=1,\cdots,L$ with $L$ being the total number
of sky map pixels. The most of elements $a(k,i)=0$ except for $a(k,i=k^+)=1$
and $a(k,i=k^-)=-1$, where $k^+$ denotes the pixel observed by the plus-horn
and $k^-$ the pixel observed by the minus-horn at an observation $k$.

The normal equation of Eq.~\ref{dt1} is
\begin{equation}
\label{ne}
\mathbf{Mt=A^Td}
\end{equation}
with $\mathbf{M=A^TA}$.

The Eq.~\ref{ne} can be expressed as
\begin{eqnarray*}
N_i^+t_i-\sum_{k^+=i}t_{k^-}-\sum_{k^-=i}t_{k^+}+N_i^-t_i
=\sum_{k^+=i}d_k-\sum_{k^-=i}d_k \\
(i=1,2,\cdots,L)~.
\end{eqnarray*}
Where $\sum_{k^+=i}$ means summing over $N_i^+$ observations while the pixel
$i$ is observed by the plus-horn and $\sum_{k^-=i}$ means summing over $N_i^-$
observations while the pixel $i$ is observed by the minus-horn,
and the total number of observations for the pixel $i$ is $N_i=N_i^++N_i^-$.
From the above equations we can derive the following iterative formula
\begin{eqnarray}
\label{mm}
t_i^{(n+1)}=\frac{1}{N_i}(\sum_{k^+=i}(d_k+t_{k^-}^{(n)})-\sum_{k^-=i}
(d_k-t_{k^+}^{(n)})) \nonumber \\
 (i=1,2,\cdots,L)~.
\end{eqnarray}

With Eq.~\ref{mm} when the number $n$ of iteration is large enough, we get
the final solution $\hat{t}_i=t_i^{(n)}$ for each pixel $i$.
The Eq.~\ref{mm} and the approximate iterative formula used by the WMAP team
\cite{hin03}, both have good performance for the differential data of a
noiseless instrument.

\subsubsection{Temperature distortion by instrument and observation imbalances}
The differential data of WMAP contain errors caused by the instrument imbalance:
the output of a WMAP radiometer from the plus-horn and mines-horn is not a purely
differential response to sky signals \cite{jar03,jar07}. Instead of the ideal case
$d_k=t_{k^+}-t_{k^-}$, a real observed differential data is
\begin{equation}
\label{imb}
 d_k=t_{k^+}-t_{k^-}+\delta_k\,,
\end{equation}
where the difference distortion
\begin{equation}
\label{delta}
\delta_k=x_{im}(t_{k^+}+t_{k^-})
\end{equation}
with the transmission imbalance factor $x_{im}$ being valued between about 0.001 and 0.02
for different bands~\cite{jar07}.

\begin{figure}
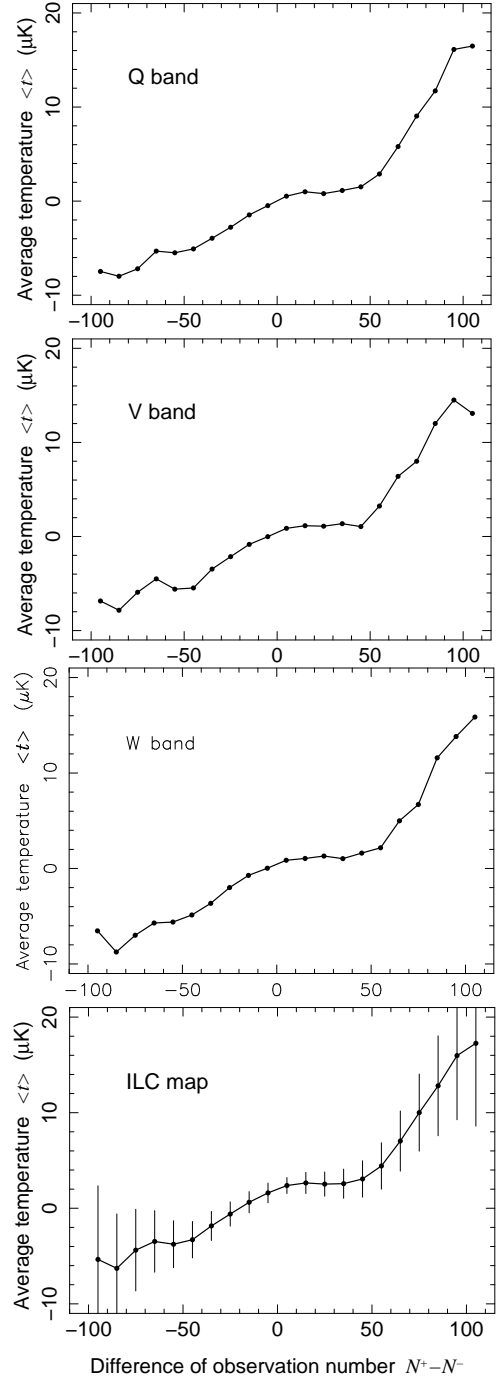

\begin{center}
\includegraphics[width=0.25\textwidth,angle=270]{f3a.ps}
\includegraphics[width=0.25\textwidth,angle=270]{f3b.ps}
\includegraphics[width=0.248\textwidth,angle=270]{f3c.ps}
\includegraphics[width=0.281\textwidth,angle=270]{f3d.ps}
\end{center}
\caption{Average temperature vs. observation number difference
$N^+-N^-$ from WMAP5 data
for Q-band, V-band, W-band and ILC maps. Error bars are only marked on the graph for
ILC map, which are similar for other graphs. }
 \label{dn}
   \end{figure}

To roughly estimate the magnitude of the transmission imbalance effect
on recovered temperatures, we use the first approximation derived from Eq.~\ref{mm}
with the initials $t^{(0)}_k=0$
\begin{equation}
\label{ti1}
t_i^{(1)}=\frac{1}{N_i}(\sum_{k^+=i}d_k-\sum_{k^-=i}d_k)\,.
\end{equation}
Substituting Eq.~\ref{imb} into Eq.~\ref{ti1} we get
\begin{equation}
\label{ti1-1}
t_i^{(1)}=\frac{1}{N_i}\{\sum_{j=1}^{N_i^+}[t_i-t_{j}+\delta_{j}]
-\sum_{j=N_i^++1}^{N_i}[t_{j}-t_i+\delta_{j}]\}\,,
\end{equation}
where $j\in R_i$, $R_i$ is the scan-ring with angular radius $141\degree$
to the pixel $i$. For $N_i^+$ observations $j=1,\cdots,N_i^+$, the plus-horn
points to the pixel $i$ and the mines-horn scans along the ring $R_i$; and
for $N_i^-$ observations $j=N_i^++1,\cdots,N_i$, the mines-horn
points to the pixel $i$ and the plus-horn scans along the ring $R_i$.
From Eq.~\ref{ti1-1} we can estimate the temperature distortion
\begin{equation}
\label{tdis}
t_i^{(1)}-t_i\simeq \frac{N_i^+-N_i^-}{N_i}\overline{\delta}_R-\overline{t}_R\,,
\end{equation}
where $\overline{t}_R=\sum_{j=1}^{N_i}t_j/N_i$,
$\overline{\delta}_R=\sum_{j=1}^{N_i}\delta_j/N_i$ and $j\in R_i$.

From  Eq.~\ref{tdis} we can see that a hot source contained on the scan-ring
$R_i$ will let  the ring average temperature $\overline{t}_R\gg 0$ and
the recovered temperature $t_i^{(1)}\ll 0$, in other words,  a hot foreground
source might systematically make the recovered  temperatures on its scan-ring lower.
We have indeed found such systematic distortion in released WMAP5 maps:
scan-rings of hot sources are significantly cooled and strongest anti-correlations
between pixel temperature and temperature of scan-ring with different separation
angle $\theta$  appear at $\theta\sim141\degree$~\cite{liu08}.

Eq.~\ref{tdis} indicates that there might exist another kind of systematic distortion
in recovered temperatures caused by transmission imbalance of radiometers
through observation imbalance $N_i^+\ne N_i^-$.
Eq.~\ref{delta} shows that the difference distortion $\delta_k$
of an observation $k$ dependents on the measured temperatures $t_{k^+}+t_{k^-}$.
Although the transmission imbalance factor $x_{im}$ is rather small, but
the difference distortion $\delta_k$ can be considerable if any one of the measured
temperatures is high enough. From  Eq.~\ref{tdis} we see that
for balance observations -- the observation numbers of two horns $N_i^+=N_i^-$,
the difference distortion can not affect the recovered temperature.
If the scan-ring $R_i$ of pixel $i$ contains hot sources and
the observations are imbalance $N_i^+\ne N_i^-$, the recovered
temperature might be distorted. To check this effect,
we arrange the temperatures $t_i$ of all pixels $i$ out of the foreground
mask KQ75 from a released WMAP5 map in order of the observation number
difference $\Delta N_i=N_i^+ - N_i^-$ from the WMAP5 TOD data.
Fig.~\ref{dn} shows the dependence of temperature averaged over a difference
interval of 10 vs. difference  $\Delta N$ between observation numbers by plus-horn
and mines-horn. For Q-, V-, W-band and ILC maps, temperature distortions
systematically changing along with $\Delta N$ as expected from above discussion
on Eq.~\ref{tdis} are evidently exhibited in Fig.~\ref{dn}.

\begin{figure}
\includegraphics[width=0.301\textwidth,angle=270]{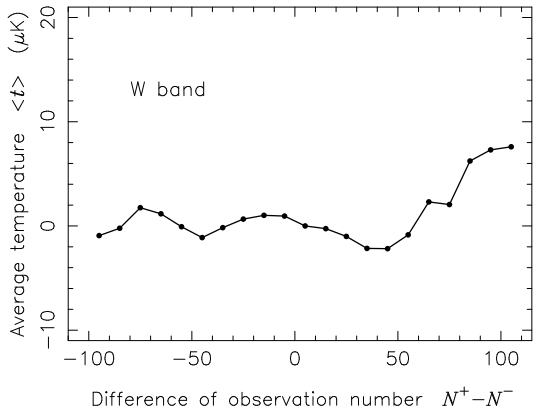}
   \caption{Average temperature vs. observation number difference $N^+-N^-$ from WMAP5
W-band map after differential imbalance compensation with a compensate
factor $\delta=50$\,$\mu$K.}
 \label{dn1}
   \end{figure}

To roughly estimate the average magnitude of difference distortion,
we try to correct the error by introducing a constant compensate factor $\delta$
for the WMAP5 differential data $d_k$ to generate
\begin{equation} 
\label{delt}
   d^*_k=d_k-\delta
\end{equation}
 and reconstruct a new temperature map from the
correlated TOD data {\bf d}$^*$. Fig.~\ref{dn1} shows the $N^+-N^-$ dependence
from the new WMAP5 W-band map with $\delta=50$ $\mu$K, where the
observation imbalance effect is much mitigated. Fig.~\ref{dn} and Fig.~\ref{dn1}
show that temperature distortions caused by the transmission imbalance of radiometers
are remarkable and scan-rings for pixels with strong imbalance observations
($\Delta N>50$) may generally contain hottest sources.

\subsubsection{$N$-$\Delta N$ correlation}
\begin{figure}
\includegraphics[width=0.301\textwidth,angle=270]{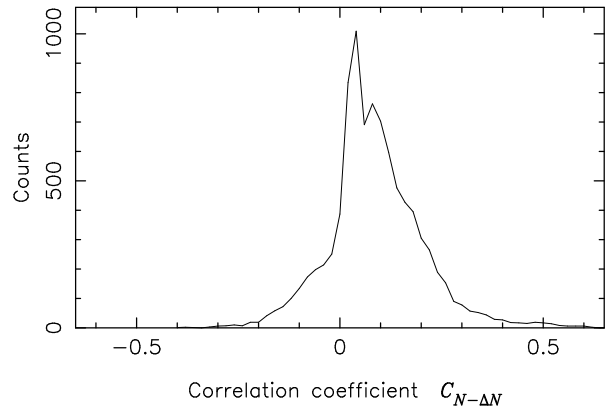}
   \caption{Histogram of $\Delta N$-$N$ correlation coefficients.}
 \label{ndn}
   \end{figure}

The dependence of average temperature $\<t\>$ vs. observation number difference $\Delta N$ 
shown in Fig.~\ref{dn} indicates that the input transmission imbalance of WMAP 
differential radiometers and unequal horn coverage ($\Delta N\ne 0$) may generate 
the $t$-$N$ correlation observed in released WMAP maps, if a considerable 
$N$-$\Delta N$ correlation exists. To check it, we calculate correlation coefficients
 $C_{_{N-\Delta N}}(i)$  from WMAP5 W1-band data  with the same procedure
we use to calculate $C_{_{t-N}}(i)$ in \S3.1: the correlation calculation
summations are taken over all WMAP pixels $j$ within a spherical cap
centered at the vertex $i$ with an angular radius of $10\degree$,
the vertexes $i$ are defined with $r5$ resolution of HEALPix pixelization scheme
in the sky sphere, if more than $20\%$ of pixels of a cap are inside the Galactic mask
KQ85 the cap is no longer used.  Fig.~\ref{ndn} shows the histogram of correlation coefficients
 $C_{_{N-\Delta N}}(i)$. From Fig.~\ref{ndn} we can see that strong correlation 
between the observation number $N_i$ and difference
$\Delta N_i=N_i^+-N_i^-$ of a sky pixel $i$
exists in observations used for making WMAP temperature maps.
For most pixels $i$,
the observation number $N_i$  is positively correlated
with the observation imbalance $\Delta N_i$,
then the temperature distortion by instrument and observation imbalances
demonstrated in this section
should contribute to the detected $t-N$ correlation presented in \S3.1
through the $N$-$\Delta N$ correlation.

\section{DISCUSSION}
Unexplained large scale anomalies, e.g. the orientation 
of large-scale patterns in respect to the ecliptic frame, 
the north-south asymmetry of temperature fluctuation power 
etc, have detected in WMAP data. 
Their origin has long puzzled scientists.  
 Several radical explanations with an
anisotropic cosmology have already been proposed.
However, it is, in particular, hard to imagine that there exists a special
direction related to the ecliptic or Galactic plane in the early universe.
In this work, we show that 
 the large-scale features of the WMAP observation number map
and observation number fluctuation distribution are very similar
to the large-scale non-Gaussian modulation feature in released 
WMAP temperature map (\S2), and find that significant correlation
between pixel temperature and observation number exists in WMAP data (\S3.1),
such correlation can be produced by the systematic effect of WMAP
instrument and observation imbalances on temperature maps (\S3.2). 
The detected observation number correlation
should contribute, at least partly, to the apparent similarity
between the large-scale features of the WMAP observation exposure
and temperature maps,
and hopefully provide a natural way to explain some large-scale
anomalies in released WMAP temperature maps.

Inhomogeneity of observation numbers used in WMAP map-making 
is emerged at different sky scales, which should generate systematic errors 
in WMAP temperature maps in a wide range of angular scale through the significant 
$t$-$N$ correlation revealed in this work. 
To limit systematic artifacts, a large amount works have been performed by 
the WMAP team. Due to the differential nature of WMAP observations,
it is a difficult task. As an example, though the effect of input transmission
imbalance from radiometer nonidealities has been noticed, calibrated and 
modified by the WMAP team \cite{jar03,jar07}, we still find remarkably systematic
dependence of temperature vs. observation number difference between the two horns  
residual in released WMAP maps. 
As shown in \S3.22, hotspots in the sky can distort recovered temperatures 
of pixels on large part of the sky with a complicated way.
It has to be pointed out that using Eq.~\ref{delt} with a constant compensation factor 
to correct the effect of horn imbalance is just to roughly estimate the average 
magnitude of differential imbalance distortion. What shown in Fig.~\ref{dn1} is only 
on the meaning of the average. 
More works have to be done to find a proper approach to modify the effect of imbalance 
differential observation to recover a corrected temperature map.

The real accuracy of cosmology parameters is the most
important issue for high precision cosmology. Systematical temperature
errors and structured noise fluctuations existed in CMB maps will certainly
distort the angular power spectrum and the best-fit
cosmology parameters as well. It is obviously needed to further study the errors
in WMAP temperature and noise fluctuation maps caused by the observation
inhomogeneity and imbalance.
The systematic distortions detected by us in released WMAP maps come from the WMAP's
differential nature. The next CMB mission Planck is
designed to measure the CMB anisotropy with completely different mode and expected
to be unaffected by such kind of distortions.

\section*{Acknowledgments}
The referees are thanked for their helpful comments and suggestions and Prof. S.N. Zhang
for suggestion on presentation of detected
non-Gaussianity of $t-N$ correlation.
This work is supported by the National Natural Science Foundation of China 
(Grant No. 10533020), the National Basic Research Program of China 
(Grant No. 2009CB-824800), and the Directional Research Project 
of the Chinese Academy of Sciences (Grant No. KJCX2-YW-T03).
The data analyzed in this work are obtained through the HEASARC on-line service
provided by the NASA/GSFC.

\label{lastpage}
\end{document}